\documentclass[prb,twocolumn,superscriptaddress,showpacs,floats,floatfix]{revtex4-1}
\usepackage{dcolumn,amsmath,latexsym,graphicx,amssymb}
\usepackage[usenames,dvipsnames]{color} \usepackage{multirow}
\usepackage{array}
\usepackage{xcolor}
\RequirePackage{ifpdf} 
\begin{document}



\title{Unraveling the Interplay between Quantum Transport and Geometrical Conformations in Monocyclic Hydrocarbons Molecular Junctions}

\author{A. Martinez-Garcia}
\affiliation{Departamento de F\'\i sica Aplicada and Instituto
Universitario de Materiales de Alicante (IUMA), Universidad de Alicante, Campus de San Vicente del Raspeig, E-03690 Alicante, Spain. }

\author{T. de Ara}
\affiliation{Departamento de F\'\i sica Aplicada and Instituto
Universitario de Materiales de Alicante (IUMA), Universidad de Alicante, Campus de San Vicente del Raspeig, E-03690 Alicante, Spain. }

\author{L. Pastor-Amat}
\affiliation{Departamento de F\'\i sica Aplicada and Instituto
Universitario de Materiales de Alicante (IUMA), Universidad de Alicante, Campus de San Vicente del Raspeig, E-03690 Alicante, Spain. }

\author{C. Untiedt}
\affiliation{Departamento de F\'\i sica Aplicada and Instituto
Universitario de Materiales de Alicante (IUMA), Universidad de Alicante, Campus de San Vicente del Raspeig, E-03690 Alicante, Spain. }

\author{E. B. Lombardi}
\affiliation{Department of Physics, Florida Science Campus,
University of South Africa, Florida Park, Johannesburg 1710, South Africa}

\author{W. Dednam}
\affiliation{Department of Physics, Florida Science Campus,
University of South Africa, Florida Park, Johannesburg 1710, South Africa}

\author{C. Sabater}
\email{Electronic mail address: \textcolor{blue}{carlos.sabater@ua.es}}
\affiliation{Departamento de F\'\i sica Aplicada and Instituto
Universitario de Materiales de Alicante (IUMA), Universidad de Alicante, Campus de San Vicente del Raspeig, E-03690 Alicante, Spain. }

\date{\today}

\begin{abstract}
In the field of molecular electronics, particularly in quantum transport studies, the orientation of molecules plays a crucial role. 
This orientation, with respect to the electrodes, can be defined through the cavity of ring-shaped monocyclic hydrocarbon molecules. In this manuscript, we unveil the geometrical conformation of these molecules when they are trapped between two atomically sharp electrodes through a combination of dynamic simulations, electronic transport calculations based on density functional theory, and break junction experiments under room conditions. Moreover,  we present a novel criterion for determining the molecular orientation of benzene, toluene, (aromatic) and cyclohexane (aliphatic) solvents. Our findings for the identification of the molecular orientations on gold metal nanocontacts and their associated transport properties, can improve the understanding of molecular electronics using more complex cyclic hydrocarbons.

\end{abstract}
\maketitle

\section{Introduction} 
All the instruments and techniques used to measure the electronic transport of atomic-sized contacts were developed over more than three decades ago\cite{Agrait1993,Krans93}. At the beginning of that period the electronic transport of atomic contacts made of  metals\cite{Krans1995},  semi-metals \cite{SbRuitenbeek}, and  superconductors\cite{Scheer1998} were studied, leading to an immediate interest in molecular junctions. The molecular electronics field initially considered very simple molecules like diatomic hydrogen and deuterium \cite{Smit2002}, which paved the way for obtaining knowledge from an experimental point of view. For various reasons and based on attempts at trying to identify exotic properties in the electronic transport field, innumerable types of different molecules have been studied \cite{Venkataraman2006, Venkataraman2022Topo, HerreElectomi2006, HerreMayor2019,Yelin2013, Taniguchi2011, Yoshida2015, tewari2019, SabaterOP3, Cuevasbook, Agrait2003, Jan2019, DeAra2022Elve}. The creation of a molecular junction basically depends on its 
 physical state and the method of deposition of the molecules, which can be grouped into three categories. The first one corresponds to the gas phase molecules, which are associated with the unique possibility of blowing the gas close to the electrodes. The second group comprises molecules in the solid state that can be deposited by thermal evaporation onto the junctions \cite{Oren2021}. The final group is composed of molecules that can be delivered in solution, and then, the usual way to capture the target molecules is to wait for the solvent to evaporate.
 
It is still widely believed that the solvent disappears or has a negligible effect on electronic transport after evaporation \cite{vanVeen2022}. However, recent works \cite{DeAra2022Elve} have demonstrated  that  the solvent's presence is indeed detectable via Scanning Tunnelling Microscopy (STM) topography and STM-Break Junction (BJ) experiments. In our previous published experimental results \cite{DeAra2022Elve}, we unmasked the molecular electronic signature  of the aromatic  (benzene, toluene) and aliphatic  (cyclohexane) molecules\cite{Quek2009, alifatic}. However, a systematic study in solvents exploring the relationship between electronic transport and the geometric orientation of the molecules has yet to be conducted.

Recent works based on atomically precise binding conformations \cite{Kamenetska22} have inspired us to clearly identify the relationship between electronic transport and molecular orientation by using Classical Molecular Dynamics (CMD) simulations and Density Functional Theory (DFT) electronic transport calculations \cite{DeAra2022Elve}. Based on our simulations, we have classified the typical final contact for benzene, cyclohexane, and toluene. Furthermore, we have calculated the electronic transport using DFT for several scenarios provided by CMD. Finally, we compare our simulations and calculations with electronic transport experiments. The good agreement of this comparison helps us to unmask the relationship between the conductance and the orientation of molecules between the gold electrodes. 

\section{Materials and methods}

\subsection{Identifying the orientation of the cyclic molecules}
\begin{figure}[h!]
 \centering
 \includegraphics[width=0.35\textwidth]{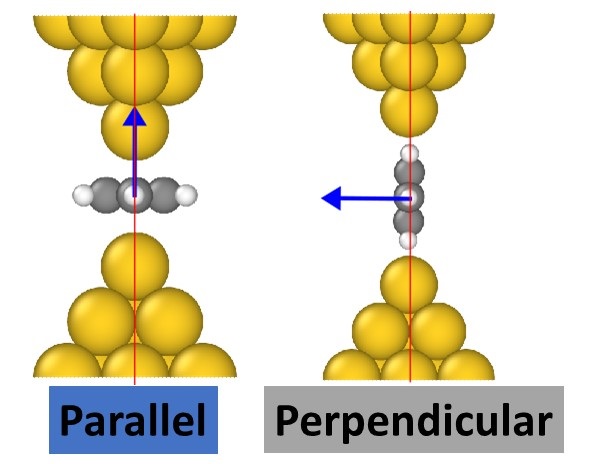} 
 \caption{Ilustration of``parallel''and ``perpendicular'' configurations of benzene molecule between two gold electrodes}
 \label{Orientation}
\end{figure}

From a mathematical point of view, the normal vector is the simplest way to identify a plane. Thanks to the fact that the aromatic solvents (such as benzene and toluene) and aliphatic solvents (such as cyclohexane) analyzed in our manuscript are monocyclic in nature, we can use their internal cavity as an orientation plane. Its normal vector can help to identify if we are in parallel or in perpendicular configuration with respect to the alignment vector defined by the electrodes. In Fig. \ref{Orientation}  blue arrows indicate the normal vector over the plane defined by the cavity of the benzene molecule. The red line indicates the direction of the vector defined by the alignment of the electrodes, in this case, dominated by the center of the apex atoms of the upper and bottom gold electrodes. As an adopted criterion we have defined parallel configuration when the normal vector is in the same direction of the alignment of the electrodes (see left illustration of Fig. \ref{Orientation}). On the other hand, when the angle formed by the normal vector with respect to the alignment vector direction is between 16 and 90 degrees, we say that we are in the perpendicular configuration  (see right illustration of Fig. \ref{Orientation}, where the extreme case of 90 degrees between both lines is depicted).

\subsection{Calculations of conductance-based on Density Functional Theory and Classical Molecular Dynamics simulations}
To calculate the conductance, we employed DFT combined with the non-equilibrium Green’s function (NEGF) approach to quantum scattering  \cite{louis2003keldysh}. The electronic transport calculations were performed using the well-established code \texttt{ANT.GAUSSIAN} \cite{palacios2001fullerene, palacios2002transport, ANTG}. This code is built on top of and interfaces with \texttt{GAUSSIAN09}. To ensure  the high quality \cite{Zotti22}  of our calculations, we used the \texttt{LANL2DZ} basis set\cite{LANL2DZ} for the atoms of the molecules and a few of the adjacent metal layers on either side of the molecules, while the outer metal layers were described by the smaller tight-binding-like \texttt{CRENBS} basis set \cite{CRENBS}.  

\begin{figure}[h]
\centering
\includegraphics[width=0.49\textwidth]{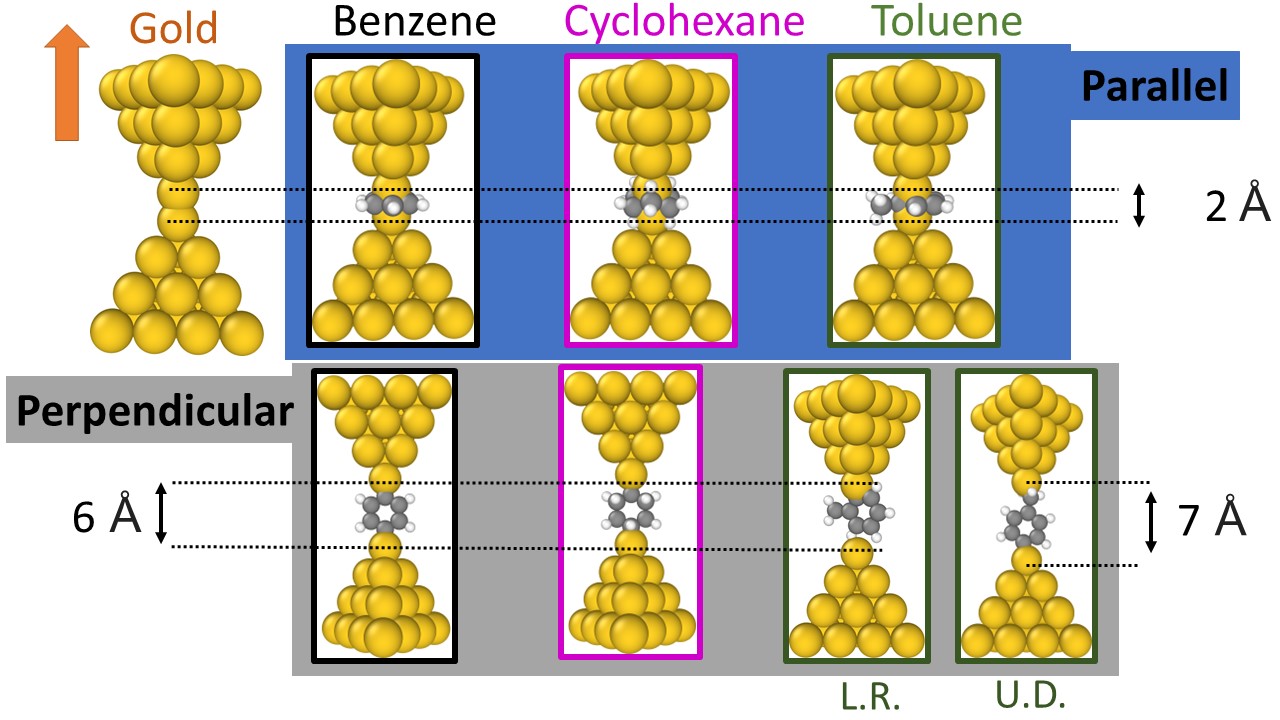} 
\caption{The upper left panel shows the type of gold point contact used. The orange arrow indicates the direction of the pull-off of the upper electrode. The blue rectangle next to it shows the initial structures of the benzene, cyclohexane and toluene junctions for one of the two configurations studied in this work. The starting distance between the two gold apex atoms of the tips in the ``parallel'' configuration is depicted with dashed lines and is $2$ \AA.  Three of the ``perpendicular'' configurations have a starting distance between the gold tip apex atoms, indicated by dashed lines, of $6$ \AA. In the fourth ``perpendicular'' configuration, the toluene molecule with the methyl group attached to the upper electrode has gold tip atoms that start apart $7$ \AA. The ``perpendicular'' configuration of the toluene is labelled as L.R. and U.D. which indicates if the methyl group is located left-right or up-down.}
\label{DFTconfiguration}
\end{figure}

In this work, we have calculated the electronic transport in two scenarios: (i) For idealized initial test structures such as those shown in Fig. \ref{DFTconfiguration} and (ii) for electrodes stretched by CMD simulations as described below. For each of  these scenarios, the DFT quantum transport methodology remained the same. 

A simulation based on classical molecular dynamics essentially solves Newton's second law in order to obtain the trajectories of all the atoms involved in the phenomena it is meant to describe \cite{allen1989computer,frenkel2002}.
We have traced  the evolution of a collection of atoms at a very fine time resolution ($\sim 1$ fs).
To simulate the rupture-formation cycles of the gold nanowire with the target molecules, we used the \texttt{LAMMPS} code\cite{plimpton1995fast,lammps2}  with a ReaxFF potential \cite{ReaxFF11,ReaxFF16}, which reproduces  quite well the mechanical  and catalytic behaviour of metal-organic compounds \cite{Soria2018,Gomzi21}.
In order to reproduce the experimental conditions, the simulations were performed at a constant temperature of 300 K with a Nosé-Hoover thermostat. 
The simulated nanowire is created with an area of narrower cross-section  ($\approx$ 17 x 17 x 25 \AA$^3$) grown along the (001) crystallographic direction \cite{RoleSab18}, where the rupture will take place. In this area of constriction, we place four molecules on each of the four square crystallographic facets of the gold nanowire (see the "Input" panels in Fig. \ref{MDsumar}); hence, sixteen molecules are deposited in total for each initial condition. This is repeated for each of Benzene, Cyclohexane, or Toluene in turn.

\subsection{STM-Break Junction experiments}
The molecular electronics experiments were performed  by  STM-BJ approach under ambient conditions. As electrodes, we used two gold wires (0.5 mm in diameter and of 99.99\% purity as supplied by Goodfellow \cite{Goodfellow}). The voltage applied in all the experiments was 100 mV. The current through our molecular junctions was converted into volts in three stages of the current-voltage amplifier with respective gains of ${10^6}$, ${10^8}$, and ${10^9}$ $V/A$, in order to be recorded by our DAC system. This homemade instrument was  tested and used in a previous publication \cite{DeAra2022Elve}.
Knowing the applied bias voltage and current, we can easily  calculate  the conductance, which is expressed in quantum of conductance ($G_{0}=2e^2/h$), where factor two comes from the spin degeneracy, $e$ is the charge of the electron and $h$ is the Planck's constant.
Historically,  the representation of  conductance vs the relative displacement of the electrodes has been called a ``traces of conductance''. These traces can be labelled   as ``rupture'' or ``formation'', depending on whether the electrodes are pulled apart or pushed together, respectively.
In this manuscript, we have only studied the rupture traces. As internal protocol, we always use ultra-clean gold electrodes, and we verify the extent of ``cleanness'' via a logarithmic histogram, which for clean electrodes is characterised by the non-existence of any peak between the atomic contact ($\approx$ 1 $G_0$) or any level of noise in the I-V converter.
Only after obtaining ultra-clean gold, the organic solvent is deposited via drop casting over electrodes \cite{DeAra2022Elve}. 

\section{Results and discussion}
\begin{figure}[h]
 \centering
  \includegraphics[width=0.45\textwidth]{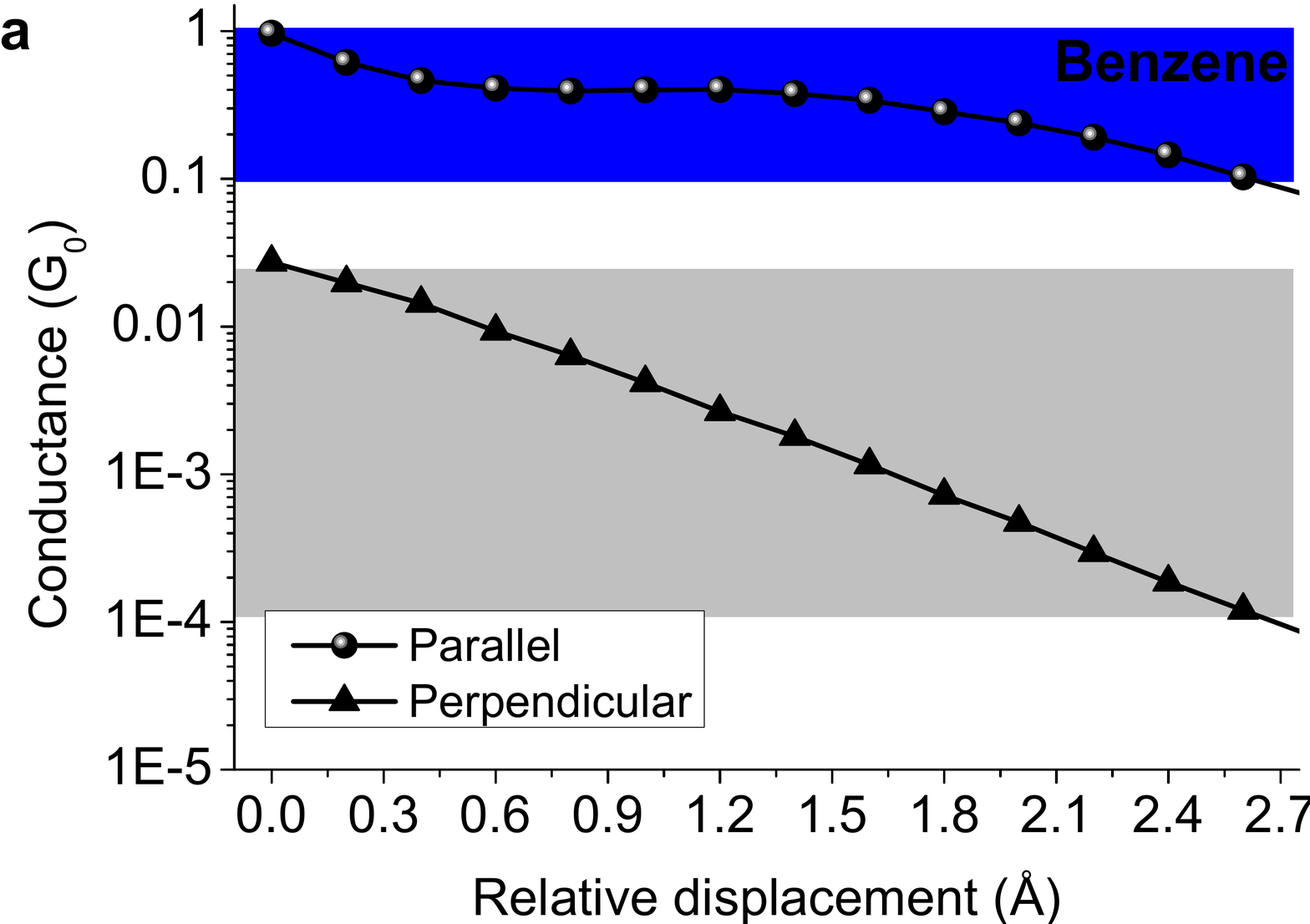}
    \includegraphics[width=0.45\textwidth]{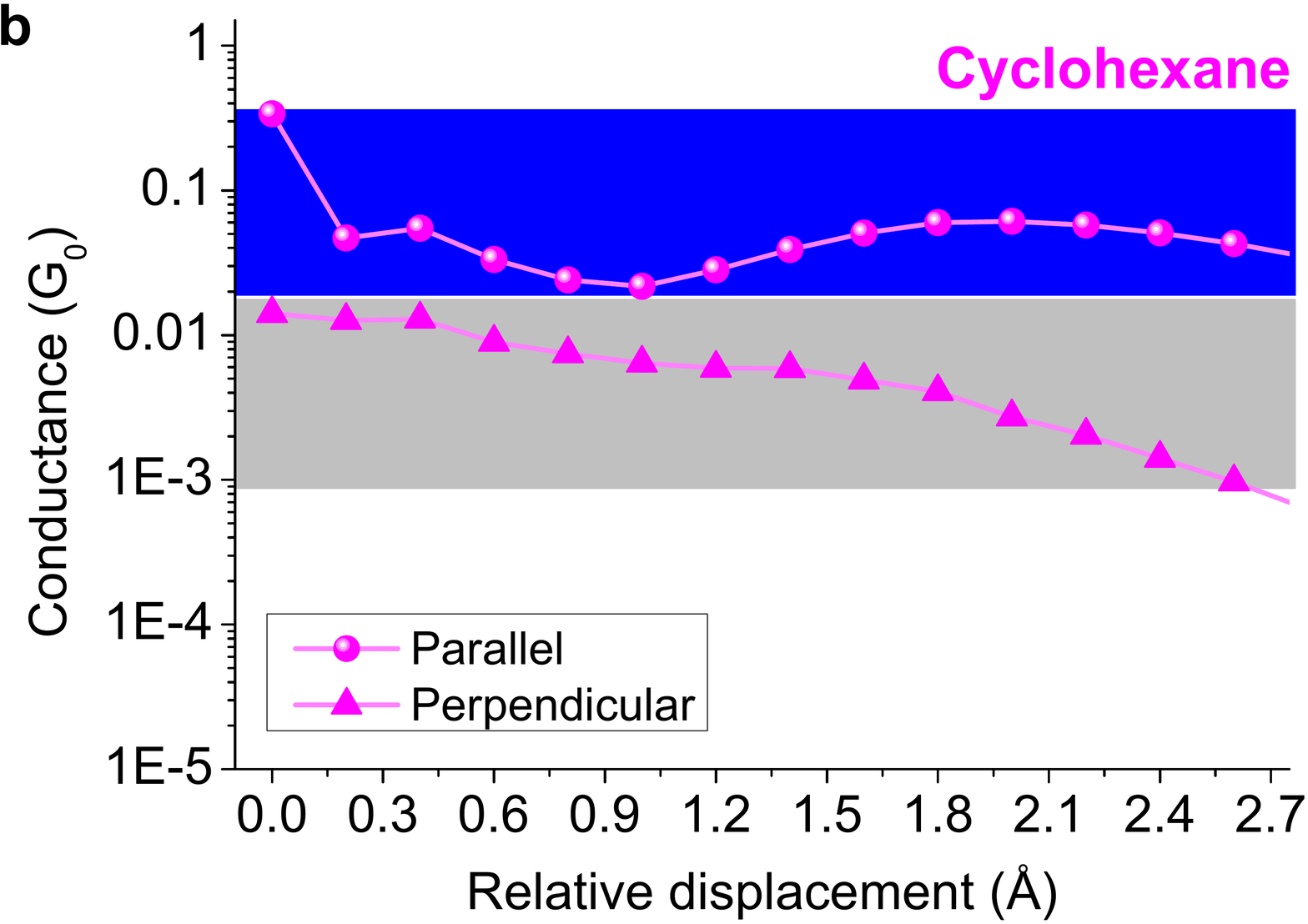}
     \includegraphics[width=0.45\textwidth]{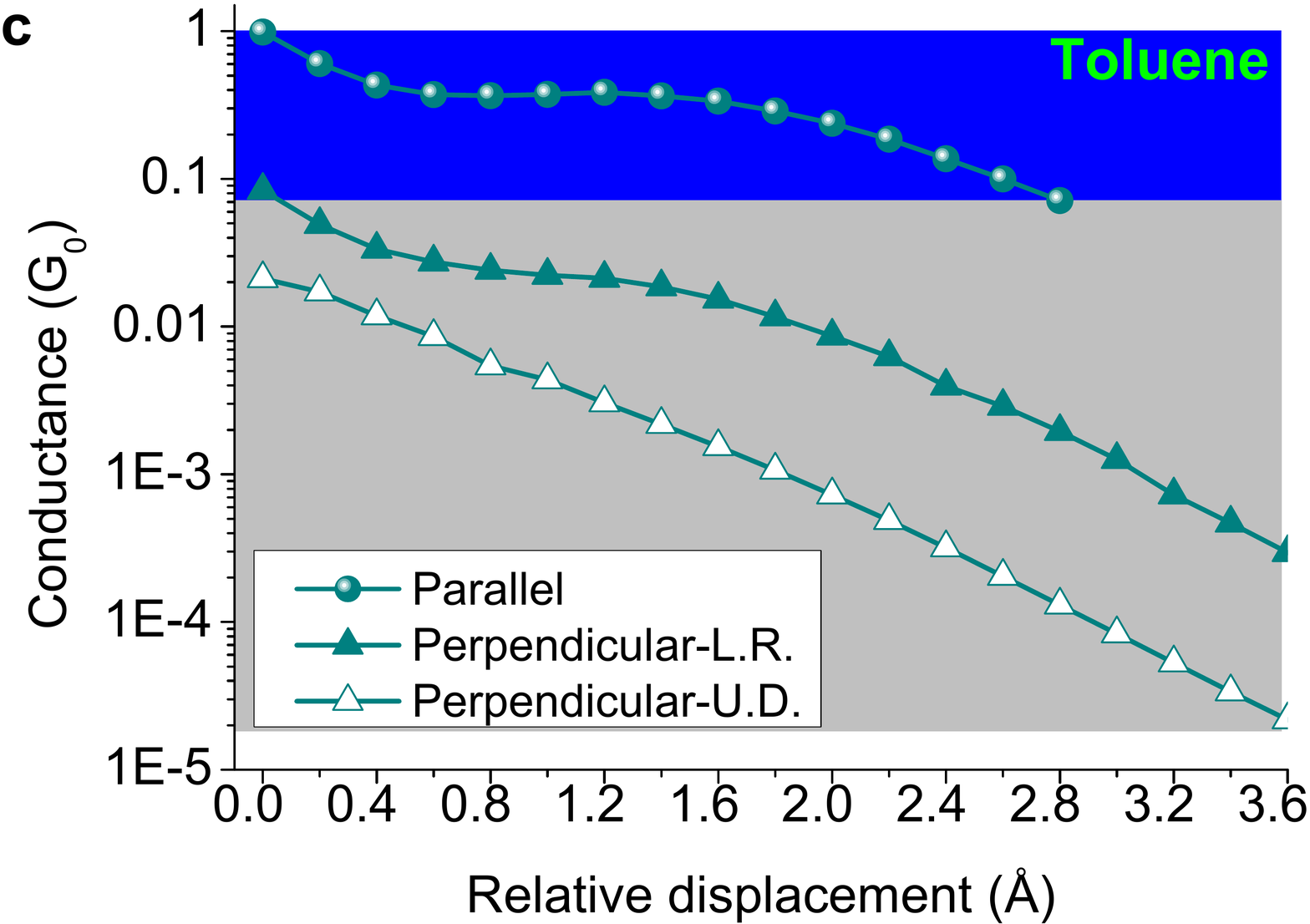}
 \caption{Calculated electronic transport vs relative displacement of the initial structures of  benzene, cyclohexane, and toluene, labelled as a,b, and c, respectively. Coloured markers represent the calculated values of benzene (black), cyclohexane (pink) and toluene (green). }
 \label{DFTpure}
\end{figure}

To study the change in conductance with relative displacement and compare the orientation of molecules between the electrode tips, we assumed that the molecule and electrode atoms are static and that the only movement allowed is the bulk displacement of the upper electrode (as shown in Fig. \ref{DFTconfiguration}, where the gold contact and its orange arrow indicating the direction of movement are present).
We have studied the evolution of the conductance vs the relative displacement of the upper electrode for benzene, cyclohexane and toluene molecules in the parallel and perpendicular configurations (see Fig. \ref{DFTpure}). In all the cases we have moved the upper electrode in step intervals of $0.2$ \AA \space without relaxation. At every step, we have calculated the electronic transport (conductance) by DFT+NEGF. In Fig. \ref{DFTpure}, the three panels show the conductance in units of $G_0$ versus displacement up to $2.5-3.6$ \AA \space (depending on the molecule). This range has been selected to exclude the tunnelling regime ($2.5-3.6$ \AA \space to $6-7$ \AA)  due to lack of information of interest (see  for full range and details sup. inf. Fig. \ref{figSMDFTfull}.).

Fig. \ref{DFTpure} upper, middle and bottom panels show the evolution of the conductance for benzene, cyclohexane and toluene, respectively.
In each panel, spheres-lines represent parallel configurations and triangles-lines perpendicular configurations. Moreover, the colour code used in Fig. \ref{DFTconfiguration} for every molecule also carries over to these panels a, b, and c. In direct correspondence with Fig. \ref{DFTconfiguration}, the ``parallel'' and ``perpendicular'' structures are indicated by blue and grey shaded areas, respectively.  The height of this rectangle indicates the maximum and minimum values attributable to each configuration. This colour code of blue and grey facilitates understanding  of the experimental data.

From Fig. \ref{DFTpure}, one of the first observations we can make is that ``parallel'' configurations of benzene and toluene  exhibit very similar conductance traces during stretching. Additionally, the calculated conductance traces for the perpendicular cases of benzene and toluene ``perpendicular-U.D.'' show the same behaviour. These results suggest that the electronic transport in these aromatic molecules is relatively similar. By contrast, the aliphatic molecule cyclohexane shows different electronic transport values of around $ \sim 10^{-3} G_{0}$ for a displacement of $2.5$ \AA. These ``toy-model'' conductance's calculations of idealized structures by DFT+NEGF are not representative of the vast quantity of possible positions that can be adopted by the molecule during the experiment. For this reason, we have decided to simulate via CMD simulations the different types of contacts that can likely be produced.

\begin{figure}[h]
 \centering
  \includegraphics[width=0.5\textwidth]{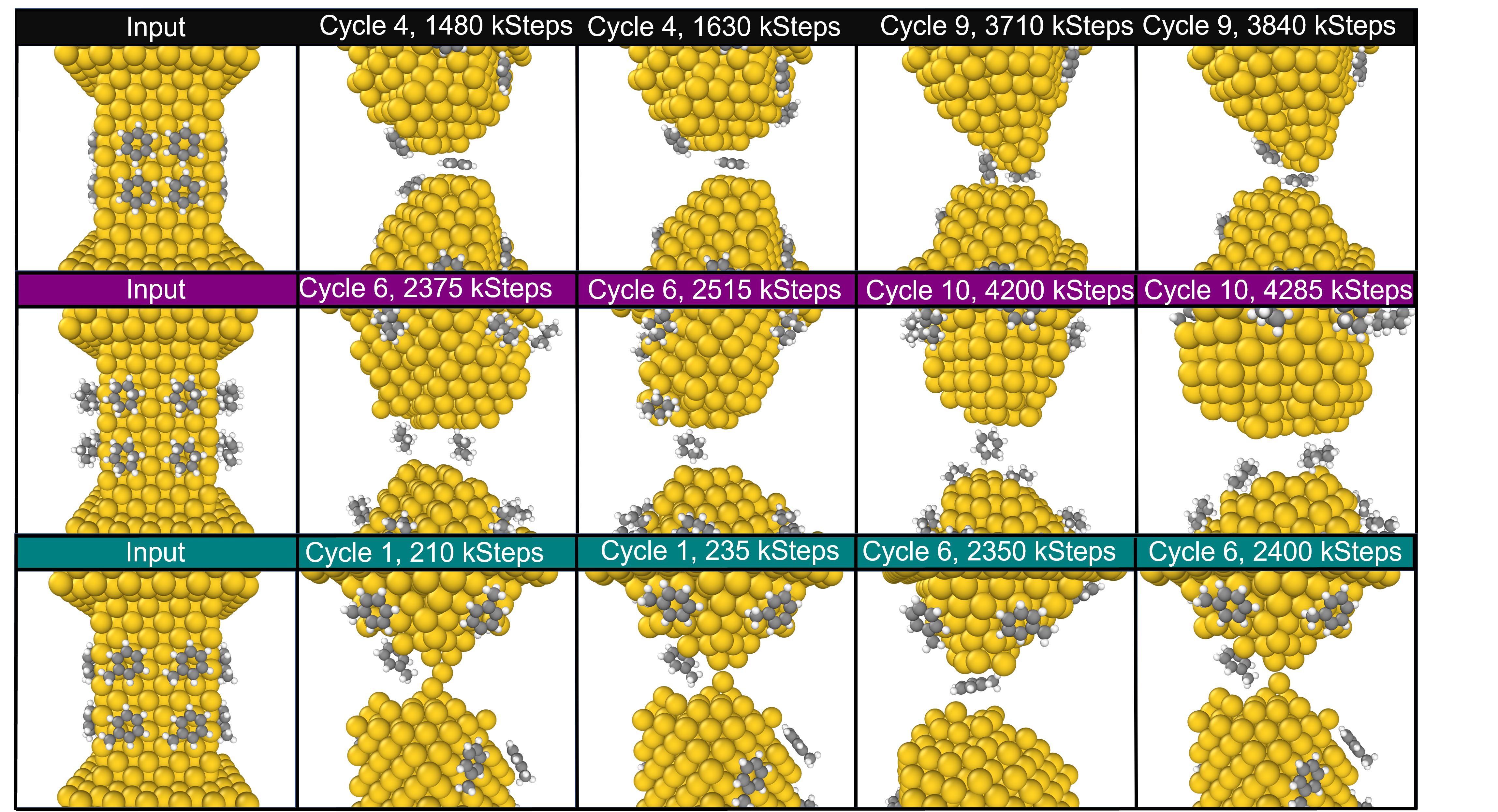}
 \caption{Snapshots of the simulation  during the process of rupture and formation cycles. Each row  corresponds to a different molecule: benzene (top), cyclohexane (middle) and toluene (bottom).}
 \label{MDsumar}
\end{figure}

Fig. \ref{MDsumar} shows a summary of the snapshot of the cycles of rupture and formation obtained by CMD simulations. Each row corresponds to a different molecule, as indicated by the colour code (black, purple, and green for benzene, cyclohexane, and toluene, respectively). At the top of each snapshot, we indicate the simulation cycle and step in units of ksteps (where 1 step equals 1 femtosecond). The first column depicts the initial input structure used. For each molecule, multiple junction rupture and formation cycles are repeated. Representative snapshots are taken from different cycles at various steps in the cycle, illustrating the different positions that the molecules can take.

 We analyzed only the rupture cycles thanks to experimental data, that only is based on rupture traces. In particular, we have focused on the last point immediately before rupture and we have created the following classification hierarchy.
\begin{figure}[h]
 \centering
  \includegraphics[width=0.49\textwidth]{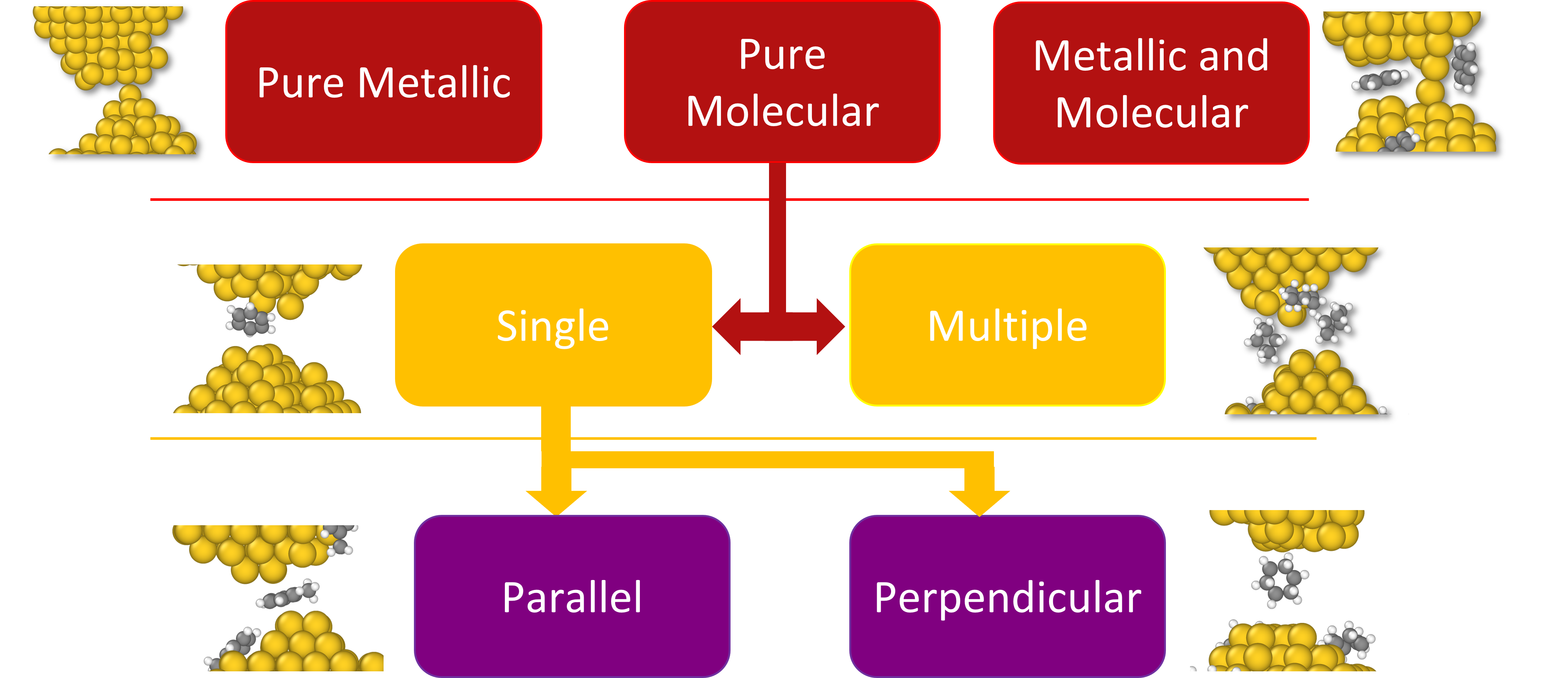}
    \caption{Classification hierarchy of molecular contacts used in this work. To the left and right, we show example structures that illustrate each type of last contact.}
 \label{Clasification}
\end{figure}

At the top of this hierarchy, the last contact before rupture can be purely metallic or purely molecular or a combination of both. The red blocks shown at the top in Fig. \ref{Clasification} represent these three common types of final molecular contacts. The structures on the left and right of the red blocks depict final contacts composed of pure metal and a mixture of metal and molecules, respectively. When the final contact is purely molecular, two possibilities can occur: either a single molecule or multiple molecules are held by the nanojunction. This secondary classification of the pure molecular last contact is shown in the second row of Fig. \ref{Clasification} and is represented by yellow blocks in the diagram. Moreover, the structures to the left and right of these yellow blocks illustrate single or multiple molecular junctions respectively. As Fig. \ref{DFTconfiguration} illustrated, a single molecule can be captured either in a ``parallel'' or a ``perpendicular'' configuration by the junction, as indicated by the purple blocks in the final and lowest level of the classification hierarchy. The structure on the left of this last row illustrates the parallel orientation, while that on the right corresponds to the perpendicular one. 

Thanks to our classification, we  can determine how many times a given type of last molecular contact occurs in our simulations (see  table \ref{tab:Cycles resume}). The first column indicates the molecule used in the simulation. The second column corresponds to the total number of rupture cycles analyzed. The third column represents the number of events in which we found a purely metallic contact at the moment of last contact. The fourth column is the number of events that we have identified as last contact involving a molecular contact (independently of whether the contacts contain single or multiple molecules). Finally, the fifth column shows the number of events in which we observed combined molecular and metallic contacts. 

\begin{table}[h]
\caption{Classification of the last type of contact during the 
 simulated process of the rupture.}
    \centering
    \begin{tabular}{|c|c|c|c|c|}
        \hline 
           {Molecule} & Cycles  & Metallic & Molecular & Met+Mol \\
        \hline 
              Benzene &  $15$ & $3$ & $4$ & $8$ \\ 
         \hline 
         Cyclohexane & $11$ & $0$ & $11$ & $0$ \\ 
     
        \hline 
         Toluene &  $16$ & $1$ & $10$ & $5$ \\ 

        \hline  
       
    \end{tabular}
    \label{tab:Cycles resume}
\end{table}

\begin{figure}[h]
 \centering
\includegraphics[width=0.49\textwidth]{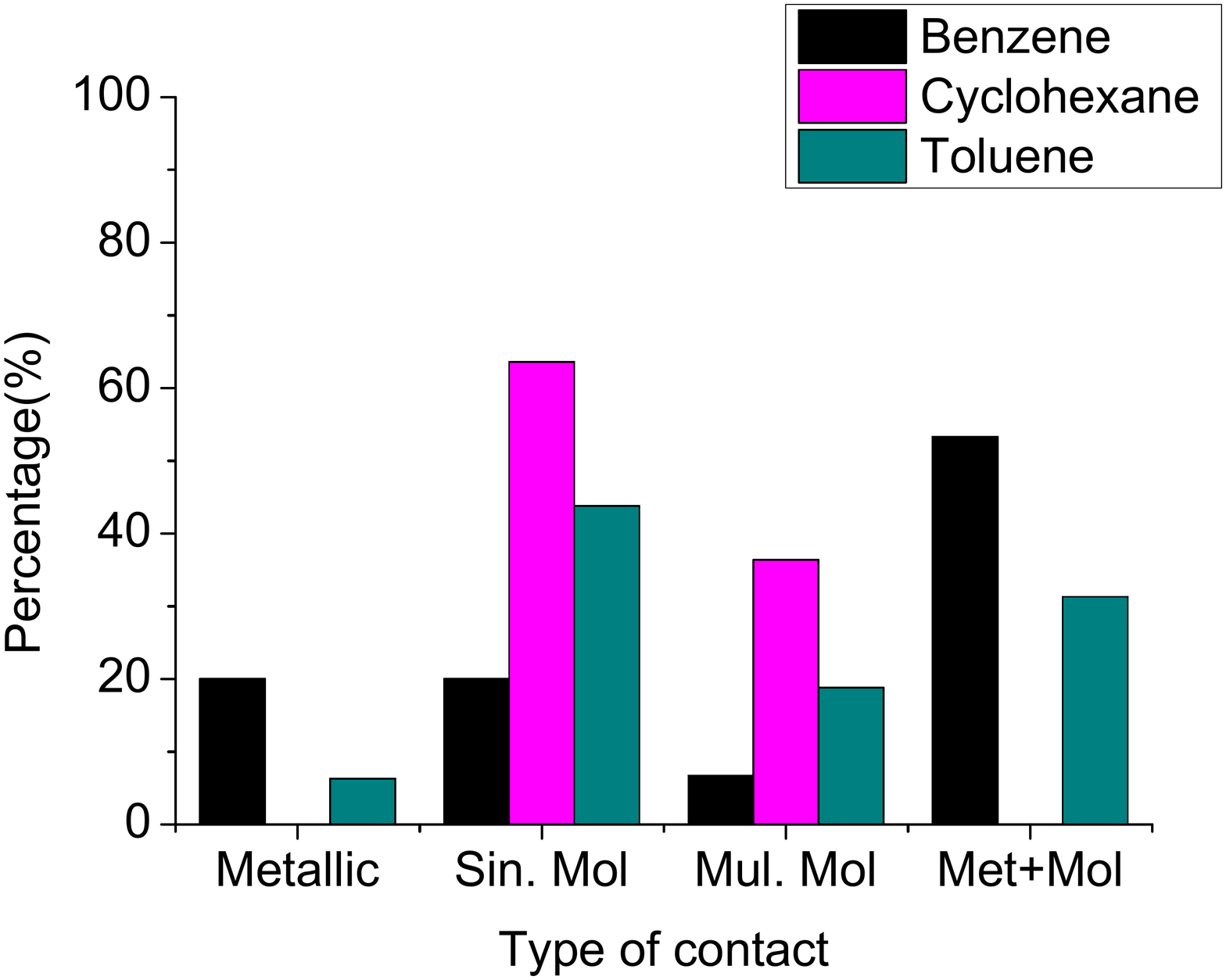} 
    \caption{Percentage  calculated from  table \ref{tab:Cycles resume} for the different types of last rupture in the benzene, cyclohexane and toluene.}
 \label{Percentage}
\end{figure}

 Fig. \ref{Percentage}  represents in a bar graph  the percentage of the events obtained in table \ref{tab:Cycles resume}, the colour code used is the same as in previous plots. From this figure, we observe  clearly that the most probable last contact for the cyclohexane and toluene  is the single-molecule junction, and for the case of benzene is the combination of the metallic-molecular junction. However, from the results shown in table \ref{tab:Cycles resume} and  Fig. \ref{Percentage}, it is difficult to deduce a refined classification of the type of pure molecular junction that is possible. For this reason, we present table \ref{tab:Cyclesrefined} where the first column indicates the molecule, the second and third columns correspond to the single parallel and single perpendicular configurations, respectively, and the fourth column corresponds to the multiple molecular junctions that can be found at the moment of last contact. 

\begin{table}[h]
\caption{Refined classification of the pure  molecular junctions of our simulation results. The number of events of Single Parallel (Sin. ara.), Single Perpendicular (Sin. Perp.), and Multiple Molecular (Mul. Mol.) junctions.}
    \centering
    \begin{tabular}{|c|c|c|c|}
        \hline 
           {Molecule}  & Sin. Para.  & Sin. Perp.  & Mul. Mol. \\
        \hline 
              Benzene & $2$ & $1$ & $1$ \\ 
        \hline 
         Cyclohexane  & $0$ & $7$ & $4$ \\ 
     
        \hline 
         Toluene & $3$ & $3$ & $4$ \\ 

        \hline  
       
    \end{tabular}
    \label{tab:Cyclesrefined}
\end{table}

\begin{figure}[ht]
\includegraphics[width=0.49\textwidth]{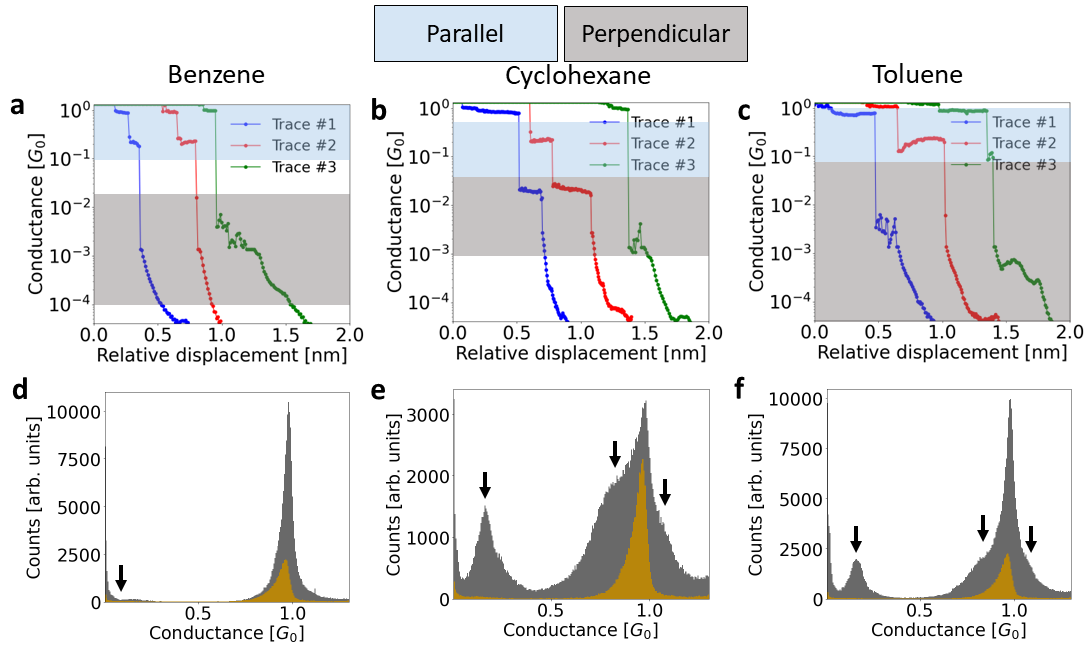} 
\caption{In panels a, b, and c, we show rupture traces for benzene, cyclohexane and toluene, respectively. Grey and blue shaded areas represent the range of values of the theoretical traces calculated via DFT based on the different geometric configurations tested. Bottom panels d,e, and f show one-dimensional experimental histograms of conductance for benzene, cyclohexane and toluene. }
\label{eXPRESULT-1Dhist}
\end{figure}

In order to ascertain the accuracy and reliability of our simulations and conductance calculations, we compare the simulation results with the experimental data. In each experiment, we first study bare gold for which we usually collect over 2000 trace files as reference. Once we have finished the analysis of pure gold, we drop cast the molecules over the bare gold and start the process of breaking and reforming of the junction, whilst at the same time collecting up to 10000 traces of conductance. Typically, molecular signatures are observed under the 1 G$_{0}$ region, so a logarithmic scale is usually most appropriate for analyzing the data statistically. A representative collection of experimental single rupture traces as a function of the stretching of the junction is shown in Fig. \ref{eXPRESULT-1Dhist} panels a, b, and c. Each trace exhibits a different initial plateau-like behaviour followed by a plateau at lower conductance values or in the tunnelling regime, related to the bridging of the electrodes by the molecules. Moreover, here we also display within the plots the blue and gray shading obtained from the DFT calculations shown in Fig. \ref{DFTpure}. These shadings help us identify if the plateau aligns with parallel or perpendicular configurations. 

When we plot conductance histograms in the linear scale, we also observe humps or shoulders around the conductance of the metallic contact (1 $G_0$). Fig. \ref{eXPRESULT-1Dhist} panels d, e, and f show the experimental histograms for each molecule in dark grey and for bare gold in dark gold colour. The signals around 1 G$_{0}$ conductance can be ascribed to two different scenarios. The first could be to a parallel configuration of the molecules compressed between the electrodes, so the coupling may be stronger and electrons hopping between the electrodes are also allowed, and hence the measured conductances are higher. The second scenario can be ascribed to the measurements of the gold atom with the molecule alongside it, providing greater stability to the junction. The data in table I and Fig. \ref{Percentage} show that the latter is the most likely scenario.

\begin{figure}[h!]
 \centering
 \includegraphics[width=0.49\textwidth]{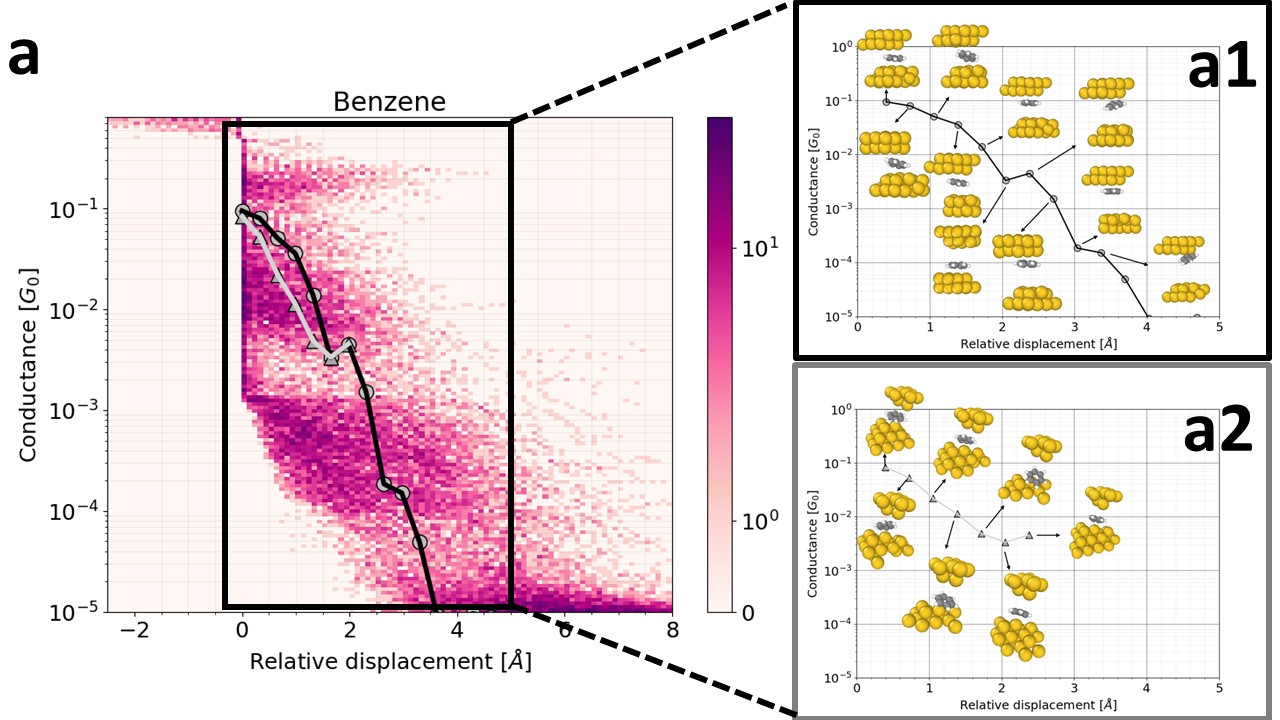}
 \includegraphics[width=0.49\textwidth]{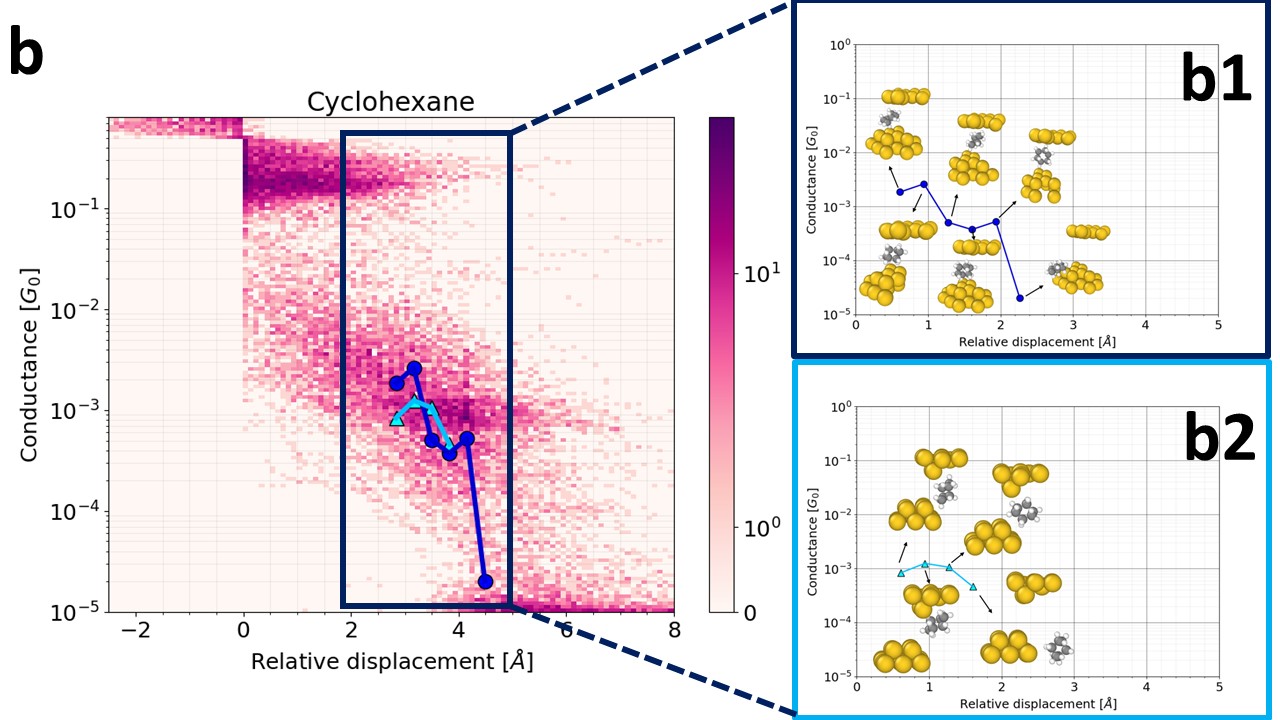} 
 \includegraphics[width=0.49\textwidth]{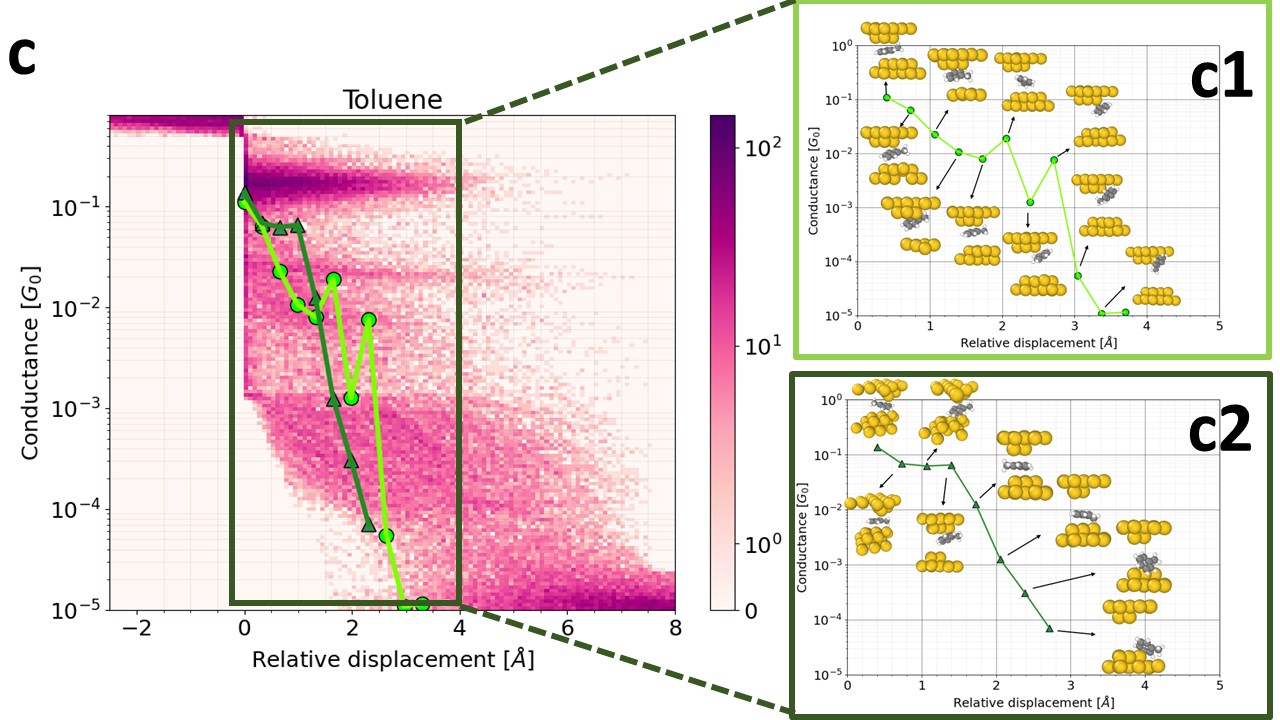}
 \caption{Experimental conductance-displacement density plots for benzene, cyclohexane, and toluene are shown in panels a, b, and c, respectively. Inside each density plot, a coloured rectangle demarcates the area where the traces of the theoretical model overlap with the experimental data. A zoom-in of the calculated traces of conductance and an illustration of the molecular contacts are shown in insets to the right of each density plot. Here, a1 and a2 correspond to benzene, b1 and b2 to cyclohexane and finally c1 and c2, to toluene.}
 \label{eXPRESULT-2Dhist}
\end{figure}
In Fig. \ref{DFTpure} idealized scenarios were shown, whose results can not reproduce accurately a dynamic process as our experiments. Hence, to identify the possible binding configurations, we have computed the conductance from some scenarios obtained via CMD, and compared it with the experimental data using 2D histograms (see Fig. \ref{eXPRESULT-2Dhist}). In constructing the experimental 2D histograms we have included only traces involving molecular junctions under 0.7 G$_{0}$, aligning them with the value of 0.5 G$_{0}$ to enhance solely the molecular contribution. Overlaid on the 2D histograms, are the computed conductance values obtained as the systems evolve, i.e. as the distance between the electrodes increases.  The colour maps on the logarithmic scale show the most repeated occurrences of the plateaus.   Rectangular frames facilitate focusing on the range where theoretical values are obtained. Two insets also accompany each panel to show the evolution of the junction with the separation distance and the different atomic arrangements of the molecules between the gold electrodes. The insets' edges are coloured in concordance with the theoretical trace. The three panels a,b, and c of Fig. \ref{eXPRESULT-2Dhist}, provide insight of the detailed molecular orientations of the molecules in the junction during the junction break process, with a high level of agreement between experimental data and the results obtained from our theoretical model consisting of CMD and DFT+NEGF.

\section{Conclusions}

Through a combination of electronic transport experiments under ambient conditions, molecular dynamics simulations and DFT transport calculations of the conductance, we have untangled the relation between electrical conductance  and binding conformations of aromatic and aliphatic molecular junctions.

 In the first set of DFT transport calculations we performed, our idealized DFT-based models reproduced the experimentally observed range of conductance versus displacement values for molecular junctions made of benzene, cyclohexane and toluene. Subsequently, classical Molecular Dynamics simulations aided in the identification and classification of the type of last contact that is likely formed immediately before rupture in the experiments. Moreover, CMD simulations generated different scenarios of the contacts that have allowed us to compute the electronic transport and obtain statistics with which to make a direct comparison with the experimental data. Consequently, the results obtained by means of the combination of CMD with DFT are in complete agreement with the experimental data, as Fig.\ref{eXPRESULT-2Dhist} shows. We further believe that the classification that we have established for the different types of molecular contacts can be applied in molecular electronics involving all manner of ring-shaped or planar molecules. 

The combination of our previous findings\cite{DeAra2022Elve} and the results presented in this manuscript reveal that the solvents benzene, cyclohexane and toluene are never fully evaporated and remain adsorbed on the electrodes. Our simulations, calculations and experiments thus allow us to obtain the characteristic conductance values of these solvents, such that when they are employed alongside other target molecules in future experiments, we will be able to clearly distinguish the conductance signatures of the solvents and their geometric relationships with the electrodes. 

\section{Acknowledgment}

This work was supported by the Generalitat Valenciana through CIDEXG/2022/45, CDEIGENT/2018/028, and PROMETEO/2021/017 and the Spanish government through PID2019-109539-GB-C41. This study forms part of the Advanced Materials program and was supported by the Spanish MCIN with funding from European Union NextGenerationEU and by Generalitat Valenciana through MFA/2022/045. The theoretical modelling was performed on the high-performance computing facilities of the University of South Africa and the University of Alicante.

\bibliography{references.bib}
\newpage
\section{Supplementary information}

\begin{figure}[h]
 \centering
 \renewcommand\thefigure{S1}  
  \includegraphics[width=0.45\textwidth]{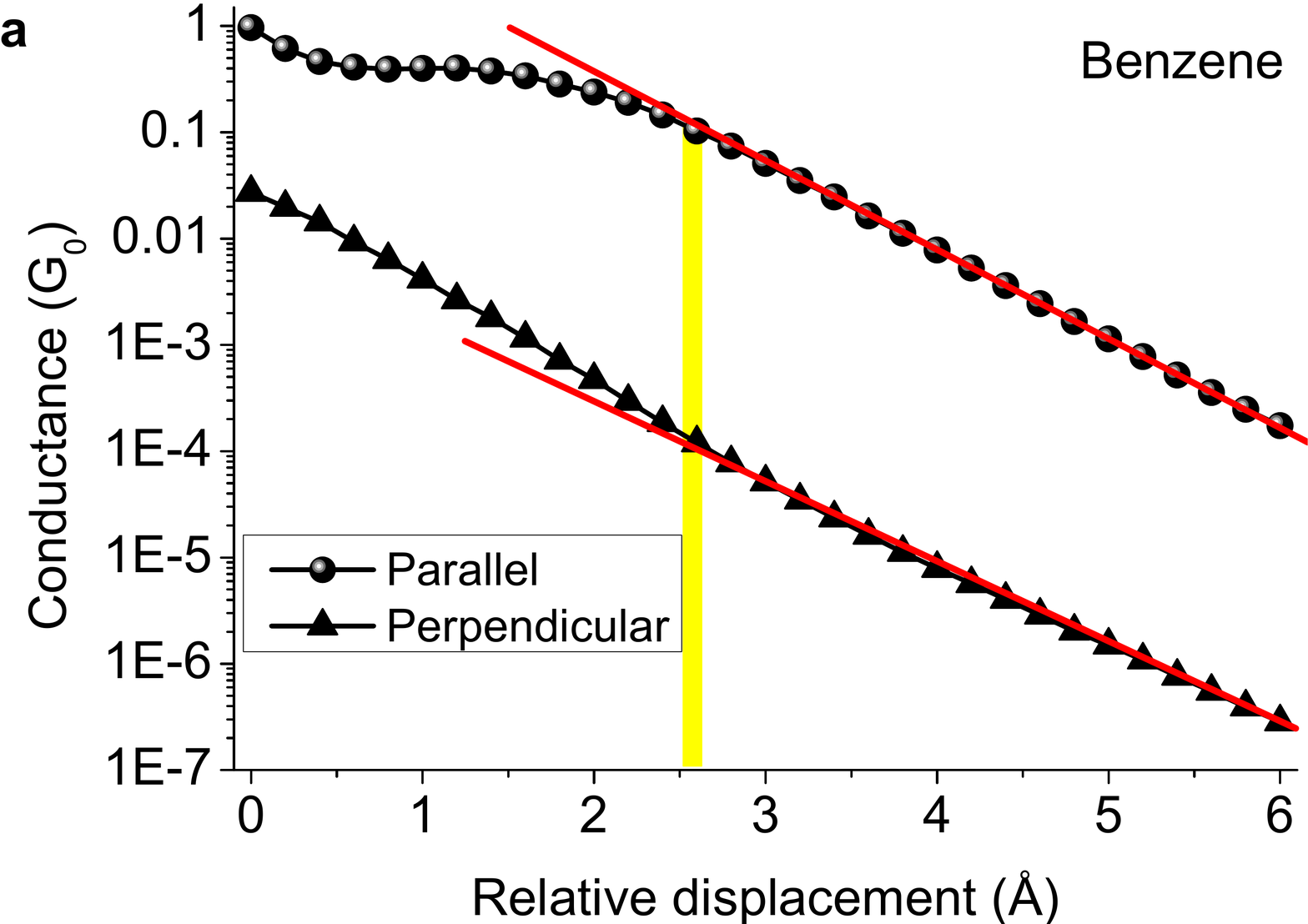}
    \includegraphics[width=0.45\textwidth]{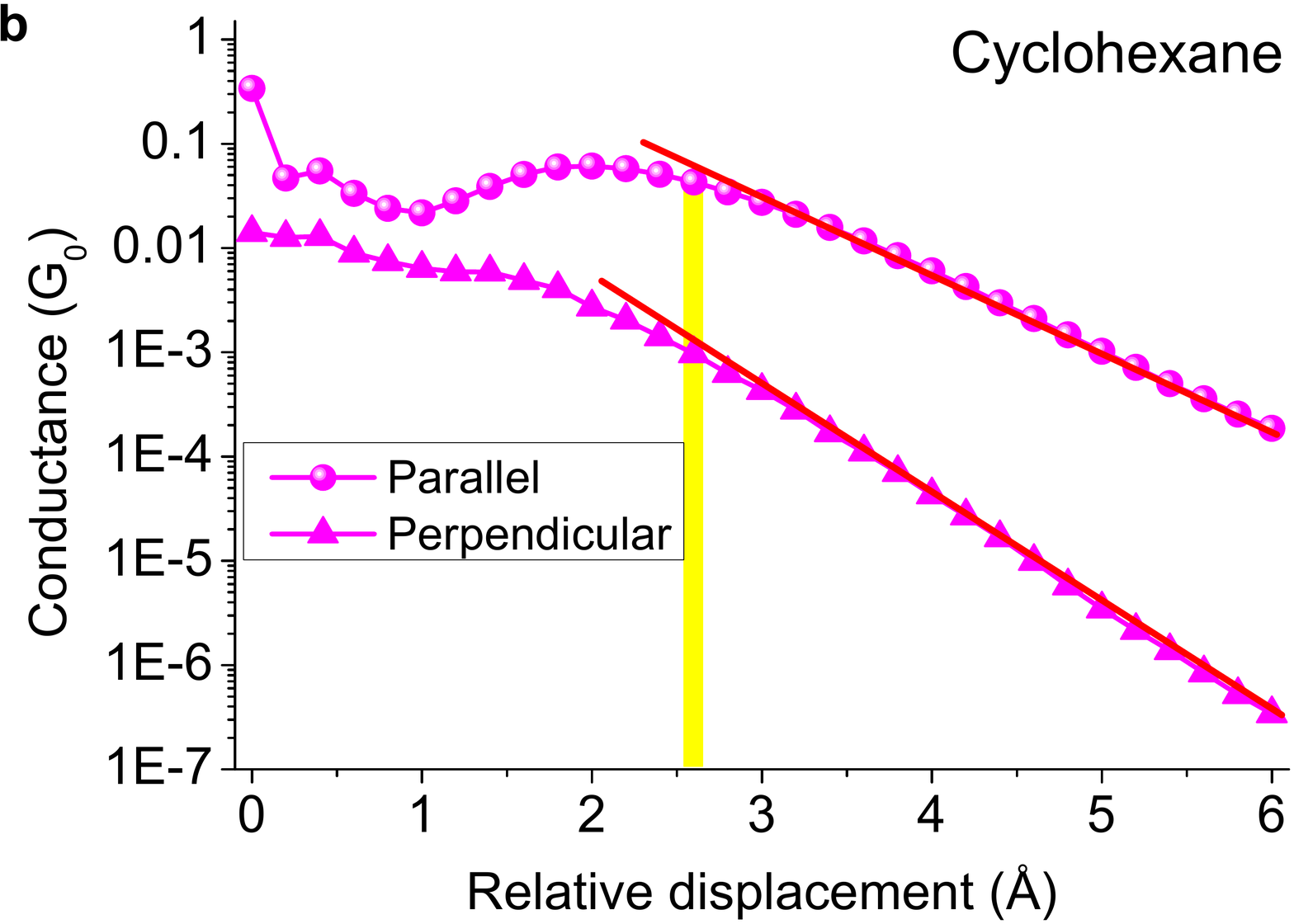}
     \includegraphics[width=0.45\textwidth]{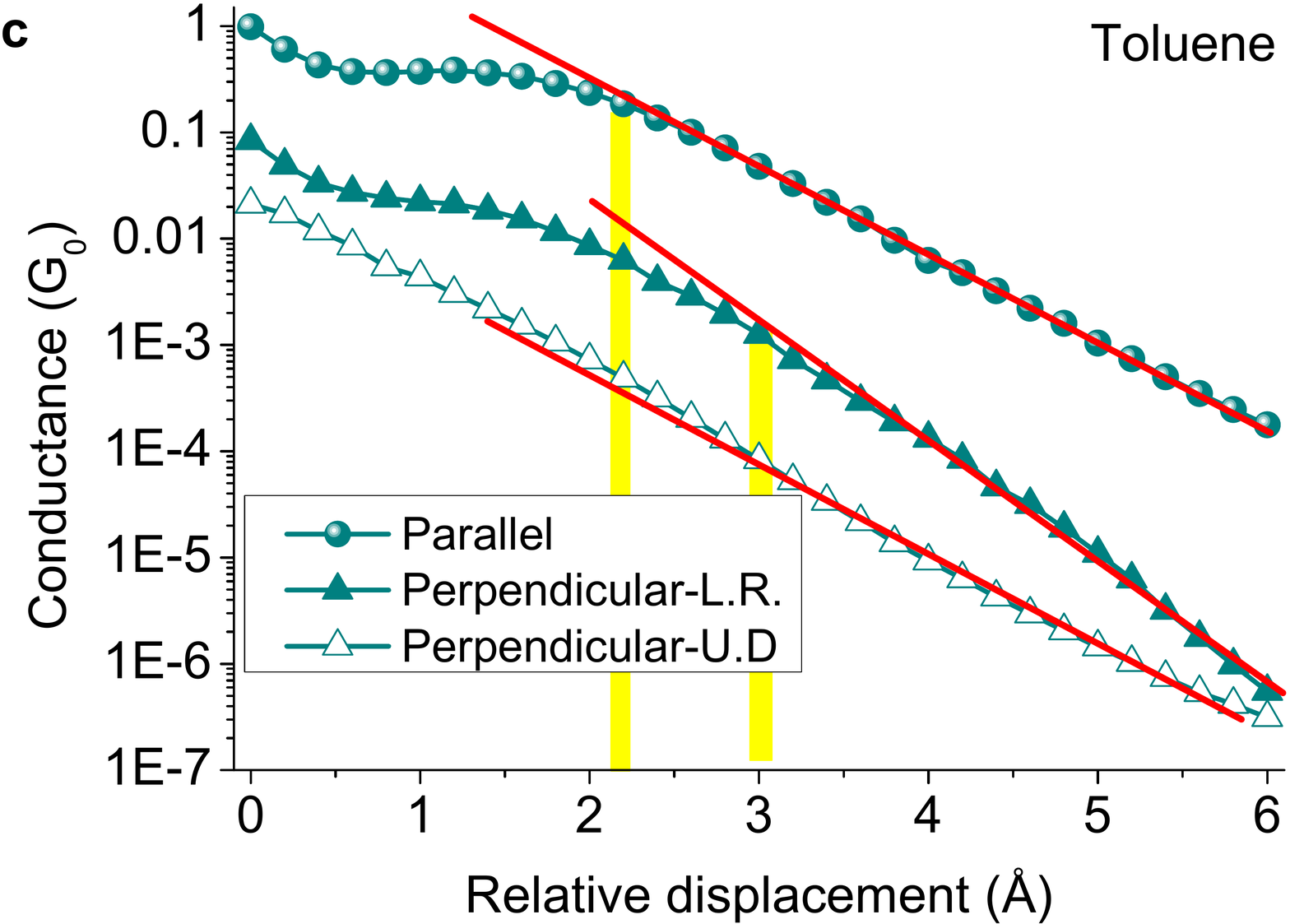}
 \caption{Electronic transport vs relative displacement calculated for the initial structures in panel (a) benzene, (b) cyclohexane and (c) toluene. Coloured markers represent the calculated values of gold (orange),  benzene (black), cyclohexane (pink) and toluene (green).}
 \label{figSMDFTfull}
\end{figure}

 Fig.\ref{figSMDFTfull}  shows the evolution of the  calculated conductance vs the relative displacement in the range $0$ \AA \space to $6$ \AA. This figure is the same as Fig. \ref{DFTpure} but without the subtraction of the tunnelling regime.   In all the cases we have moved the upper electrode in step intervals of $0.2$ \AA \space without relaxation and we  have calculated the  conductance by DFT+NEGF. The three panels show the conductance in units of $G_0$ for benzene, cyclohexane and toluene molecules in parallel and perpendicular configurations. The red lines indicate the slope of the tunnelling regime. Yellow areas highlight the first value that stat to be separated from the red trend line. We used the value of the relative displacement highlighted to create Fig. \ref{DFTpure}.

\end{document}